\documentclass[preprint,showpacs,preprintnumbers,amsmath,amssymb]{revtex4-1}

\usepackage{graphicx}
\usepackage{rotating}
\usepackage{amsmath}
\usepackage{color}
\usepackage{enumerate}
\usepackage{multirow}
\usepackage{dcolumn}  
\usepackage{booktabs} 
\usepackage{float}

\newcommand{\br}{{\bf r}}
\newcommand{\bx}{{\bf x}}
\newcommand{\be}{\begin{equation}}
\newcommand{\ee}{\end{equation}}
\newcommand{\bea}{\begin{eqnarray}}
\newcommand{\eea}{\end{eqnarray}}
\newcommand{\hH}{\hat{H}}
\newcommand{\hh}{\hat{h}}
\newcommand{\prt}{\partial_t}
\newcommand{\on}{{1 \ldots n}}
\newcommand{\onp}{{1' \ldots n'}}

\begin{document}
\title{Challenges in Truncating the Hierarchy of Time-Dependent Reduced Density Matrices Equations: Open Problems}

\author{Ali \surname{Akbari}$^\dagger$}
\thanks{These authors have contributed equally in this work.}
\author{Mohammad Javad \surname{Hashemi}$^\ddagger$}
\thanks{These authors have contributed equally in this work.}
\author{Risto M. \surname{Nieminen}$^\ddagger$}
\author{Robert van \surname{Leeuwen}$^\S$}
\author{Angel \surname{Rubio}$^\dagger$}
\affiliation{
   $^\dagger$ Nano-Bio Spectroscopy Group and ETSF Scientific Development Centre, 
	Departamento de F\'isica de Materiales, Centro de F\'isica de Materiales CSIC-UPV/EHU-MPC and DIPC, 
	Universidad del Pa\'is Vasco UPV/EHU, Av. Tolosa 72, E-20018 San Sebasti\'an, Spain.\\
   $^\ddagger$ Department of Applied Physics, Aalto University, PO Box 11100, FI-00076 Aalto, Finland. \\
   $^\S$ Department of Physics, Nanoscience Center, University of Jyv\"{a}skyl\"{a}, FIN 40014 Jyv\"{a}skyl\"{a}, Finland.}


\begin{abstract}
In this work, we analyze the Born, Bogoliubov, Green, Kirkwood and Yvon (BBGKY) hierarchy of equations 
for describing the full time-evolution of a many-body fermionic system in terms of its reduced density
matrices (at all orders). We  provide an exhaustive study of the challenges and open problems linked to
the truncation of such hierarchy of equations to make them practically applicable. We restrict
our analysis to the coupled evolution of the one- and two-body reduced density matrices, where higher
order correlation effects are embodied into the approximation used to close the equations. We prove
that within this approach, the number of electrons and total energy are conserved, regardless of the
employed approximation. Further, we demonstrate that although most of the truncation schemes
available in the literature give acceptable ground state energy, when applied to describe driven
electron dynamics exhibit undesirable and unphysical behavior, e.g., violation and even divergence 
in local electronic density, both in weakly- and strongly-correlated regimes.
We illustrate and analyze these problems within the few-site Hubbard model. The model can be
solved exactly and provides a unique reference for our detailed study of electron dynamics for
different values of interaction, different initial conditions, and large set of approximations considered
here. Moreover, we study the role of compatibility between two hierarchical equations, and positive-semidefiniteness
of reduced density matrices in instability of electron dynamics. We show that
even if the used approximation holds the compatibility, electron dynamics can still diverge when
positive-definitiveness is violated. We propose some partial solutions of such problem and point the main paths
for future work in order to make this approach applicable for the description of the correlated electron dynamics 
in complex systems.
\end{abstract}
\maketitle

\section{Introduction}
The description of dynamical processes in many-electron systems brought out of equilibrium requires a proper description of
static and dynamical correlation effects. Theoretical study of many fundamental processes of interest such as attosecond dynamics, 
high-intensity laser phenomena, light-induced phase transitions, harmonic generations, etc., require a proper theoretical framework which are 
both accurate and tractable. 

Time-dependent Hartree-Fock (TDHF) theory \cite{frenkel, dirac, thouless} was among the first methods
 devised to simulate dynamics of many-body systems. More advanced methods such as Kadanoff-Baym 
equations \cite{kadanoff, petri} and the Keldysh technique \cite{keldysh} for the Green's function
enable us to study non-equilibrium phenomena more accurately, once an approximation for the electron self-energy is chosen properly.
They are, however, computationally very demanding \cite{stan,thygesen}.

Alternatively, time-dependent density functional theory (TDDFT) \cite{hardy, ullrich} is becoming a popular option for its  performance,
accuracy and potential scalability. It describes electron dynamics in terms of the electronic density 
by mapping an interacting system to an auxiliary noninteracting system (Kohn-Sham system) with an effective potential that produces
the same time-dependent density.
This is ideal for massively parallel implementations as in its time-dependent Kohn-Sham (KS)
formulation the evolution of each of the KS orbitals is nearly independent of the others. However, lack of proper exchange-correlation (xc) functionals for 
time-dependent systems hamper its applicability. Most of the time-dependent approximations that are used so far, 
are adiabatic extensions of DFT approximations that disregards the non-locality and memory dependence of the time-dependent 
xc potential.

In this situation, reduced density matrix (RDM) theories offer a promising framework to deal with time-dependent phenomena. 
The equation of motion for each RDM can be straightforwardly derived from
the time-dependent Schr\"{o}dinger equation which contains the corresponding and one order higher RDM. The whole set of these interrelated equations
form the so-called BBGKY hierarchy since a basically similar hierarchy was initially invented and developed by Born, Bogoliubov, Green,
Kirkwood and Yvon \cite{born,bogoliubov,kirkwood,yvon} in classical statistical mechanics.
 
As it is not possible to confront the whole hierarchy, one must truncate it at some level. For instance, if one approximates  
two-body RDM in terms of one-body RDM in the first equation, 
one arrives at the time-dependent version of reduced density matrix functional theory (TD-RDMFT). 
Similar to the TDDFT, most of the approximations used in TD-RDMFT are adiabatic extensions of the existing ground state ones \cite{Muller,buijse,gritsenko}; 
and even though they can successfully describe the ground state of some strongly correlated systems \cite{Sangeeta,Ari},
they suffer from flaws such as lack of memory  when we apply them in time domain.
As an example, assuming that these approximations have even the right form, the sign adopted in each 
term generally becomes a time-dependent phase when we extend them over the time domain. Ignoring this fact by using fixed signs 
may cause problems such as time-independent occupation numbers \cite{Heiko}.
Furthermore, majority of these approximations do not necessarily conserve total energy of a system.

Some of these deficiencies will be cured if we consider propagating the first two equations of the hierarchy by approximating the three-body RDM. 
However, this prove to be a nontrivial task and in fact there are earlier attempts in nuclear dynamics \cite{Schmitt_springer,Gherega1993166} 
which show that these coupled equations can violate the inequalities related to probabilistic interpretation of RDMs for fermions.
For example, the eigenvalues of one-body RDM must remain, in principle, between zero and one \cite{coleman} but in practice they 
observed violation of these bounds, showing the non-fermionic nature of the corresponding RDM. Such behaviors were unexpected and 
it was claimed to be related to the violation of the relations between different orders of reduced density matrices.

In this paper, we study in detail the performance of such approach for different truncation schemes and show that the truncated set of equations 
may lead to instability and in many cases even divergence (in electronic density, occupation numbers, etc.). We mention the specific properties 
of approximations that are responsible for these unphysical results. We will show that lack of properties such as positive-semidefiniteness also plays a crucial role in this failure. In addition, this study prompts one to be aware of the same issues 
which may arise in building approximations in TD-RDMFT.  The article is divided into three sections. In next section, we give the theoretical background for the BBGKY hierarchy and different approximations to truncate it. In Section \ref{sec_res} we present the results and analyze the phenomenon using different approximations, and finally we conclude the work in Section \ref{conclusion}.

\section{Theoretical Background}
\label{sec1}

\subsection{N-Particle System and Reduced Density Matrix: Definitions and properties}
First in order to have a compact notation, we introduce two collections of space-spin coordinates as
\be
X_n \equiv (\bx_1 \ldots \bx_n) \quad ;  \quad \breve{X}_n \equiv (\bx_{n+1} \ldots \bx_N).
\ee
In this notation, $\Phi (X_N,t)$ denotes the normalized wave function of the system. Here, we note that throughout 
this work we employ atomic units. 

Now, we consider a system with $N$ identical particles described by the time-dependent Hamiltonian
\be
\hat{H} (t) = \sum_{i=1}^N \hat{h}_i + \frac{1}{2} \sum_{i \neq j}^N U_{ij} .
\ee
The last term describes the two-particle interactions $U_{ij} \equiv U (\bx_i \bx_j)$
where $\bx$ includes both 
space coordinates, $\br$, and spin coordinates, $\sigma$, of the particles. Usually $U$ is spin-independent
and has the form $U(\bx \, \bx') = w(|\br-\br'|)$.
The one-body part, $\hat{h}_i \equiv \hat{h} (\bx_i,t)$, will be time-dependent and of the form
\be
h (\bx,t) = - \frac{1}{2} \nabla^2 + v(\bx,t)
\ee
where $v$ is a general time-dependent external field. We can 
define the n-body reduced density matrix, $\Gamma^{(n)} $, of such system as  
\be
\Gamma^{(n)} (X_n,X_n', t) = \frac{N!}{(N-n)!} 
\int d\breve{X}_n \,\,  \Phi (X_n, \breve{X}_n,t) \, \Phi^* (X_n', \breve{X}_n,t) 
\label{nRDM}
\ee
where $d\breve{X}_n \equiv d\bx_{n+1} \ldots d\bx_N$ and $\int d\bx = \sum_\sigma \int d\br$.

Based on the above definition, several important properties of RDMs follow. First, different levels of 
RDMs are connected to each other by  
\be
\int d \bx_{n+1} \, \Gamma^{(n+1)} ( X_n \bx_{n+1} , X_n'  \bx_{n+1}, t)
= (N-n) \, \Gamma^{(n)} ( X_n , X_n', t) 
\label{trace_rel}
\ee
which we refer to it as \textit{partial trace relation}. Consequently, if $\Gamma^{(n)}$ is available, all 
RDMs with lower order can be calculated straightforwardly. 

Another important feature is that Eq.(\ref{nRDM}) implies all RDMs are positive-semidefinite which refers
to the fact that all eigenvalues of RDMs are always equal to or greater than zero. For a given order RDM, 
these eigenvalues and their corresponding eigenvectors can be calculated as 
\be
\int dX_n' \, \Gamma^{(n)} (X_n, X'_n,t) \, g_{i} (X_n',t)  =
\lambda_i (t) \, g_{i} ( X_n, t), \quad \lambda_i (t) \ge 0.
\label{eigenvalue}
\ee
Conventionally, the eigenvectors and eigenvalues of $\Gamma^{(1)}$ are called natural orbitals and natural 
orbital occupation numbers, and of $\Gamma^{(2)}$ are called geminals and geminal occupation numbers, respectively.

Moreover in the case of fermionic particles, the Pauli exclusion principle enforces  natural orbital occupation
numbers to be less than or equal to one \cite{coleman}. Thus, they have to remain between zero and one. This 
corresponds to the \textit{fermionic inequality}. 

\subsection{Equation of Motion, BBGKY Hierarchy and Conservation Laws}
\label{EOM}

Provided the initial state of the system is given, its time evolution is completely described by
the time-dependent Schr\"odinger equation (TDSE). Nonetheless, the solution of the TDSE is not 
usually possible, except for systems with few particles, due to the many degrees of freedom of the 
system. In contrast, most quantities of interest are $n$-body observables which can be 
obtained from the $n$-body reduced density matrix, $\Gamma^{(n)}$, 
and in most systems of relevance it holds that $n \ll N$. It is therefore natural to  
derive the equations of motion for reduced density matrices.
Thus, using Eq.(\ref{nRDM}) together with the TDSE, we get  
\bea
&& (i \, \prt  - \hH_\on + \hH_\onp ) \,
\Gamma^{(n)}  ( X_n, X'_n,t) = 
\nonumber \\
&&  \sum_{i=1}^n \int d\bx_{n+1}  (U(\bx_i \bx_{n+1}) - U(\bx'_i \bx_{n+1})) 
\Gamma^{(n+1)} ( X_n  \bx_{n+1}, X_n'\bx_{n+1},t)
\label{BBGKY}
\eea
where we defined
\be
\hH_\on = \sum_{i=1}^n \hh_i  + \frac{1}{2}\sum_{i \neq j}^n U_{ij}.
\ee
The entire set of $N$ equations for reduced density matrices forms  the well-known BBGKY hierarchy. 
The explicit form of the first two equations are
\be
(i  \, \prt  - \hat{h}_1 + \hat{h}_{1'} ) \, \gamma(\bx_1, \bx'_1,t) = 
\int d\bx_2 \, (U(\bx_1\bx_2) - U(\bx'_1\bx_2)) \Gamma(\bx_1 \bx_2,\bx'_1\bx_2,t)
\label{1rdm_BBGKY}
\ee
and
\bea
&& (i  \, \prt  - \hat{H}_{12} + \hat{H}_{1'2'} ) \, \Gamma(\bx_1 \bx_2, \bx'_1 \bx'_2,t) = \nonumber \\
&& \int d\bx_3 \, (U(\bx_1\bx_3) + U(\bx_2 \bx_3)- U(\bx'_1\bx_3)- U(\bx'_2 \bx_3))
\Gamma^{(3)} (\bx_1 \bx_2 \bx_3,\bx'_1 \bx'_2 \bx_3,t)
\label{2rdm_BBGKY}
\eea
where $\gamma \equiv \Gamma^{(1)}$ and $\Gamma \equiv \Gamma^{(2)}$. As it is customary in the literature \cite{bonitz},  
we call the right-hand side of Eq.(\ref{2rdm_BBGKY}) the three-body collision integral and use $S$ to refer to it. In general, 
each equation of the hierarchy relates a given-order of RDM, $\Gamma^{(n)}$, to one order higher RDM,
$\Gamma^{(n+1)}$. In order to make the BBGKY hierarchy practical, we must truncate it at some level, $n$,
by reconstructing the $\Gamma^{(n+1)}$ as a functional of lower-order RDMs. Although such reconstruction in terms of the one-body 
time-dependent density matrix is, in principle, conceivable by virtue of the Runge-Gross theorem \cite{hardy,TDRDMFT}, there
is no practical method available to find the exact functional and we have to use different approximations. 

In this work, however, we will only propagate $\gamma$ and $\Gamma $ since they are sufficient to calculate 
the dynamics of all one- and two-body observables. For instance for the case of total energy, the expectation
value of the one- and two-body part of the Hamiltonian, $E_1 (t)$ and $E_2 (t)$, can be written as
\be
E_1 (t) = \sum_i^N \langle \hat{h}_i \rangle = \int d\bx' \, h(\bx',t) \gamma (\bx,\bx',t) |_{\bx=\bx'} 
\label{e1}
\ee
and
\be
E_2 (t) = \frac{1}{2} \sum_{i,j}^N \langle U_{ij} \rangle = \frac{1}{2} \int dX_2 \, U (X_2) \Gamma (X_2 , X_2,t).
\label{e2}
\ee
To this end, we need to truncate Eq.(\ref{2rdm_BBGKY}) by approximating $\Gamma^{(3)}$ in terms of $\gamma$ 
and $\Gamma$. Several of these approximations will be discussed shortly.

At this point we highlight an important property of BBGKY hierarchy and the effect of truncation on it.  
As a direct outcome of  Eq.(\ref{trace_rel}), different levels of the hierarchy 
are compatible; namely, equations in the 
higher levels of the hierarchy are reducible to the lower-level ones. 
We refer to this link between equations as \textit{compatibility} condition that preferably should be fulfilled by a
good approximation. Thus, compatibility signifies that the highest equation is equivalent to the whole BBGKY 
hierarchy. This is not surprising since the highest equation is  
basically the original Schr\"{o}dinger equation. However, when we truncate the hierarchy by introducing
an approximation for $\Gamma^{(3)}$, the partial trace relation between $\Gamma^{(3)}$ and $\Gamma$ does not necessarily hold and thus 
it generally breaks the compatibility between Eqs.(\ref{1rdm_BBGKY}) and (\ref{2rdm_BBGKY})(see subsection \ref{compNOposit} for some exceptions). 
Consequently, when we truncate the BBGKY hierarchy, we have two generally distinct options to propagate
the equations which should be equivalent if the truncation approximation satisfies compatibility:

\begin{enumerate}[i)]
\item \textit{Propagating two coupled equations}: We can evolve both $\gamma$ and $\Gamma$ by solving
  Eqs.(\ref{1rdm_BBGKY}) and (\ref{2rdm_BBGKY}) together as coupled equations since the two equations
  most likely are not compatible anymore after approximating $\Gamma^{(3)} $ in Eq.(\ref{2rdm_BBGKY}).

\item \textit{Propagating only second equation}: To avoid the problem of compatibility between two equations, we can 
evolve only Eq.(\ref{2rdm_BBGKY}). Then we assign $\gamma$ to be the partial trace of $\Gamma$ and denote it as
$\gamma_{_{\Gamma}}$  to distinguish it from general $\gamma$. It mathematically reads
\be
\gamma_{_{\Gamma}} (\bx_1,\bx_1',t) = \frac{1}{N-1} \int d\bx_2 \, \Gamma (\bx_1 \bx_2, \bx_1' \bx_2,t).
\label{gamma_comp}
\ee
In this way, we prevent the complication of dealing with two coupled equations. 
                 
\end{enumerate}

The difference between these two approaches lies in distinction of $\gamma$ and $\gamma_{_{\Gamma}}$.
To see this, we derive the equation of motion for Eq.(\ref{gamma_comp}) by using Eq.(\ref{2rdm_BBGKY}). 
We have
\bea
i \partial_t \gamma_{_{\Gamma}} (\bx_1, \bx_1',t) &=& \frac{1}{N-1} \int d\bx_2 \, i\partial_t \Gamma (X_2, X'_2,t) |_{\bx_2=\bx_2'} \nonumber \\
&=& \frac{1}{N-1} \int d\bx_2 \, (H_{12}-H_{1'2'}) \Gamma (X_2, X'_2,t) |_{\bx_2=\bx_2'} \nonumber \\
&& + \frac{1}{N-1} \int d\bx_2 d\bx_3 (U (\bx_1 \bx_3) - U (\bx_1 \bx_3) ) \Gamma^{(3)}_{app} (\bx_1 \bx_2 \bx_3,\bx_1' \bx_2 \bx_3,t)
\label{step1}
\eea
where the approximated $3$-body RDM, $\Gamma^{(3)}_{app}$, does not necessarily integrate to 
$(N-2) \, \Gamma$ as we pointed out earlier. Now, using the explicit form of $H_{12}$, we can rewrite Eq.(\ref{step1}) as
\bea
i \partial_t \gamma_{_{\Gamma}} (\bx_1, \bx_1',t) &=& (h_1-h_{1'}) \gamma_{_{\Gamma}} (\bx_1, \bx_1',t) \nonumber \\
&& + \int d\bx_2 (U (\bx_1 \bx_2) - U (\bx_1' \bx_2)) \, \bar{\Gamma} (X_2, X_2,t) \nonumber \\
&& - \frac{1}{2} \int d\bx_2 \, (\nabla_2^2- \nabla_2^{'2}) \Gamma (X_2, X'_2,t) |_{\bx_2=\bx_2'}
\label{step2}
\eea
where
\be
\bar{\Gamma} (X_2, X'_2,t) = \frac{1}{N-1} \left[\Gamma (X_2, X'_2,t) +
 \int d\bx_3 \, \Gamma^{(3)}_{app} (\bx_1 \bx_2 \bx_3,\bx_1' \bx_2' \bx_3,t) \right].
\label{Gbar}
\ee
The last term in Eq.(\ref{step2}) can be written as a total divergence as follows
\be
(\nabla_2^2- \nabla_2^{'2}) \Gamma (X_2, X'_2,t) |_{\bx_2=\bx_2'} =
\nabla_2 \cdot \left[ ( \nabla_2 - \nabla_2')  \Gamma (X_2, X'_2,t) |_{\bx_2=\bx_2'} \right]
\ee
and hence vanishes after integration over $\bx_2$. Finally Eq.(\ref{step2}) becomes 
\be
(i  \, \prt  - \hat{h}_1 + \hat{h}_{1'} ) \, \gamma_{_{\Gamma}}(\bx_1, \bx'_1,t) = 
\int d\bx_2 \, (U(\bx_1\bx_2) - U(\bx'_1\bx_2)) \bar{\Gamma}(\bx_1 \bx_2,\bx'_1\bx_2,t).
\label{1rdm_BBGKY2}
\ee
Obviously, the equations of motion for $\gamma_{_{\Gamma}}$ and $\gamma$, Eq.(\ref{1rdm_BBGKY2}) and 
Eq.(\ref{1rdm_BBGKY}) respectively, will be equivalent if $\bar{\Gamma}=\Gamma$.
This derivation shows that Eqs.(\ref{1rdm_BBGKY}) and (\ref{1rdm_BBGKY2}), and therefore the two above-mentioned approaches, 
are generally different as a consequence of the fact that $\Gamma^{(3)}_{app}$
does not necessarily integrate to $(N-2) \, \Gamma$.

It is \textit{a priori} not clear which of the two approaches is preferable. For that, let us examine if
the approximate equations maintain important properties such as particle number and energy conservation.

\subsubsection{Particle number conservation}
First, we discuss the particle number conservation. To do this, we note that the diagonal of one-body RDM, $\gamma (\bx,\bx,t)$,
gives the particle density, $n(\bx,t)$. Now, particle number conservation means that the total number of particles, 
$N(t) = \int d\bx \, n(\bx,t)$, is independent of time, i.e. $\partial_t N(t)=0$.
This is guaranteed once the continuity equation holds. 
In fact, this is the case for both approaches since from either Eq.(\ref{1rdm_BBGKY}) or (\ref{1rdm_BBGKY2}), 
we arrive to 
\bea
\prt n(\bx,t) &=& \prt \gamma (\bx,\bx',t) \bigr|_{\bx'=\bx} \, =
-\frac{1}{2 i} (\nabla^2 - \nabla'^2) \gamma(\bx,\bx',t)  \bigr|_{\bx'=\bx} \nonumber \\
 &=& \,-\frac{1}{2i} ( \nabla^2+ \nabla \cdot \nabla'- \nabla \cdot \nabla' - \nabla'^2) \gamma(\bx,\bx',t)  \bigr|_{\bx'=\bx} \nonumber \\
 &=& \, -  \nabla \cdot \left[ \frac{1}{2i} ( \nabla - \nabla' ) \gamma(\bx,\bx',t)  \bigr|_{\bx'=\bx} \right]  \nonumber \\
 &=& - \nabla \cdot \mathbf{j} (\bx,t)
\label{continuity}
\eea
where we defined the current density, $\mathbf{j} (\bx,t)$,  as the quantity inside the braces.
This follows immediately from of the fact that the right-hand side of
both Eqs.(\ref{1rdm_BBGKY}) and (\ref{1rdm_BBGKY2}) vanishes for $\bx_1=\bx_1'$ independent
of $\Gamma$ or $\bar{\Gamma}$. Thus, the total number of particles is preserved even if we cut the
hierarchy at the first level. 

\subsubsection{Energy conservation}
Now we turn to the energy conservation. The total energy of the system is the sum of Eqs.(\ref{e1}) and (\ref{e2}). 
For the second approach where we use only Eq.(\ref{2rdm_BBGKY}), the time derivative of the total energy after 
some algebra (see Appendix \ref{appen-energy}) arrives at
\bea
\label{energy_conv}
\frac{dE}{dt} &=& \int d \bx \, \partial_t v(\bx,t) n(\bx,t) \\
&& + \frac{1}{2i} \int d\bx_1 d\bx_2 \, \nabla_1 
U (\bx_1 \bx_2 ) \cdot \left[ ( \nabla_1 -\nabla_{1'}) (\Gamma (\bx_1 \bx_2 , \bx_1' \bx_2,t)-\bar{\Gamma} 
(\bx_1 \bx_2 , \bx_1' \bx_2,t))  |_{1=1'} \right]. \nonumber 
\eea
In the absence of time-dependent potentials, $v$, this equation gives $dE/dt=0$, provided $\Gamma=\bar{\Gamma}$.
Nonetheless as we discussed, this is not generally valid in the second approach and it means the total energy
is not conserved there. In contrast, when we solve both Eqs.(\ref{1rdm_BBGKY}) and (\ref{2rdm_BBGKY}) together,
the second term of Eq.(\ref{energy_conv}) vanishes automatically (Eq.(\ref{energy_conv_first_approach}) in Appendix
\ref{appen-energy}) \cite{Wang,bonitz} and the energy remains constant. 

The energy conservation gives us a strong motivation for preferring the first approach, in which we propagate both 
equations simultaneously, over the second and this is what we do in the remainder of this work. However, 
we look into the other case only briefly in section \ref{sec_only} and show that the energy can fluctuate relatively large in 
time for all the approximations.

\subsection{Hierarchy Truncation Methods: Approximating the $\Gamma^{(3)}$}
\label{truncation}
In this section we discuss in detail different truncation schemes that we have evaluated in the present work. 
One systematic way of building these approximations is called \textit{cluster expansion} which is a method 
of reconstructing higher-order RDMs as anti-symmetrized products of lower-order ones plus a residual correlation
function\cite{Wang,Cassing_ZPhysA_1987,Cassing_ZPhysA_1988,Cassing_ZPhysA_1990,Cassing_ZPhysA_1992}.
To have a compact notation, first we define the wedge product as the anti-symmetrized product of $p$- and $m$-point functions by
\bea
&& a(X_p, X'_p) \wedge b(\breve{X}_p,\breve{X}'_p) =
\nonumber \\
&& 
\Bigl(\frac{1}{N!}\Bigr)^2 \sum_{\alpha,\beta} \epsilon(\alpha) \,  \epsilon(\beta) \, 
a(\bx_{\alpha_1}\ldots \bx_{\alpha_p}, \bx'_{\beta_1}\ldots \bx'_{\beta_p})\, 
b(\bx_{\alpha_{p+1}} \ldots \bx_{\alpha_N},\bx'_{\beta_{p+1}} \ldots \bx'_{\beta_N}).
\label{Grassman}
\eea
Here, $N=p+m$, $\alpha$ represents all permutations of the unprimed coordinates,
$\beta$ represents all permutations of the primed ones, and
the function $\epsilon(\alpha)$ returns $+1$ when the permutation $\alpha$ contains an even number of transpositions
and $-1$ for an odd number of transpositions \cite{Grassman}. For instance, the wedge product of two general 
one-particle matrices is 
\bea
a(\bx_1,\bx'_1) \wedge b(\bx_2,\bx'_2) & = & \frac{1}{4} \Bigl\{a(\bx_1,\bx'_1) b(\bx_2,\bx'_2) -  a(\bx_1,\bx'_2) b(\bx_2,\bx'_1) + \\  \nonumber 
&&  \phantom{\frac{1}{4} \Bigl\{}  a(\bx_2,\bx'_2) b(\bx_1,\bx'_1) -a(\bx_2,\bx'_1) b(\bx_1,\bx'_2) \Bigl\}.
\nonumber
\eea

Now, we illustrate the cluster expansion by some examples. The first term of the expansion of $\Gamma^{(n)}$ has 
the same form as in the noninteracting-particle picture, namely, it is an $n$-dimensional determinant of $\gamma$, with $\gamma (\bx_i,\bx_j',t)$ placed in row $i$ and column $j$.
For instance, for $\Gamma^{(2)}$, the first term reads
\bea
\left| 
\begin{array}{cc} 
\gamma (\bx_1, \bx'_1,t)  & \gamma (\bx_1 ,\bx'_2,t)  \\
\gamma (\bx_2 ,\bx'_1,t)  & \gamma (\bx_2 ,\bx'_2,t)  \\
\end{array} \right| 
\equiv 2  \, \gamma \wedge \gamma.
\label{g2_HF}
\eea
Now, we define a two-body correlation function, $\Delta^{(2)}$, as a means of the deviation of $\Gamma$ from the noninteracting form 
such that
\be
\Gamma (X_2 ,X'_2,t) = 2  \, \gamma \wedge \gamma
+ \Delta^{(2)} (X_2 ,X'_2,t).
\label{cumu1}
\ee
If we, for instance, approximate $\Gamma_{app}=2 \gamma \wedge \gamma$ and replace it in the first equation of 
the BBGKY hierarchy (\ref{1rdm_BBGKY}), we recover immediately the well-known TDHF equation. 

For $\Gamma^{(3)}$ accordingly, we use a noninteracting particle form and add anti-symmetrized products of
 $\gamma$ with the correlation function $\Delta^{(2)}$ -- that partly describe the $3$-body correlation --  plus a
remainder, $\Delta^{(3)}$, i.e.
\bea
\label{cumu2}
\Gamma^{(3)}(X_3 ,X'_3,t)
 = && \, \left| \begin{array}{ccc} \gamma (\bx_1,\bx_1',t) & \gamma (\bx_1,\bx_2',t) & \gamma (\bx_1,\bx_3',t) \\
\gamma (\bx_2,\bx_1',t) & \gamma (\bx_2,\bx_2',t) & \gamma (\bx_2,\bx_3',t)  \\
\gamma (\bx_3,\bx_1',t) & \gamma (\bx_3,\bx_2',t) & \gamma (\bx_3,\bx_3',t) \end{array} \right|  \\
&& + \sum_{i,j=1}^3 (-1)^{i+j} \, \gamma (\bx_i,\bx_j',t) \Delta^{(2)} ( \breve{\bx}_i, \breve{\bx}_j',t) 
+ \Delta^{(3)} (X_3 ,X'_3,t). \nonumber
\eea
In the second term on the right-hand side, $\breve{\bx}_j$ denotes the pair of variables in the set $(\bx_1 \bx_2 \bx_3)$ 
complementary to $\bx_j$ keeping the order of the arguments fixed; the same goes for the primed coordinates. For
example, $\breve{\bx}_2 = (\bx_1 \bx_3)$. Using the wedge product notation, we can rewrite Eq.(\ref{cumu2}) as
\bea
\label{cumu2-wedge}
\Gamma^{(3)} = 6 \,  \gamma \wedge \gamma \wedge \gamma + 9 \, \gamma \wedge  \Delta^{(2)} + \Delta^{(3)}
= -12 \, \gamma \wedge \gamma \wedge \gamma + 9 \, \gamma \wedge \Gamma + \Delta^{(3)}
\eea
in which we replaced the $\Delta^{(2)} = \Gamma - 2  \, \gamma \wedge \gamma$ from Eq.(\ref{cumu1}). Similarly, we can write the expansion for higher-order RDMs. 

The same method has been used in the Contracted Schr\"{o}dinger Equation formalism (the hierarchical set of equations for density matrices
 derived from the time-independent Schr\"{o}dinger equation) and referred to as \textit{cumulant expansion} 
\cite{Valdemoro, Nakatsuji, Mazziotti_approx, Mazziotti_CPL}. Nakatsuji and Yasuda made the expansion more grounded by deriving it using 
the relation between RDMs and Green's functions \cite{Nakatsuji}. Based on these, We are now ready to discuss a number of approximations for $\Gamma^{(3)}$:

\begin{enumerate}[i)]
\item \textit{Three-body collision-integral-free (3b-CIF) approximation.} The simplest one rises from the assumption of $\Gamma^{(3)}=0$, which removes the whole 
right-hand-side of Eq.(\ref{1rdm_BBGKY}).

\item \textit{Three-body-noninteracting approximation (3b-NIA).} This is obtained only by considering the noninteracting term of Eq.(\ref{cumu2-wedge})
\be
\Gamma^{(3)}_{3b-NIA} = 6 \,  \gamma \wedge \gamma \wedge \gamma.
\label{NIA_app}
\ee
This gives $\Gamma^{(3)}$ as a functional of $\gamma$.  

\item \textit{WC approximation.} We can, of course, climb to the next level and take also the second term of Eq.(\ref{cumu2-wedge}) into account which leads us to 
\be
\Gamma^{(3)}_{WC} = -12 \, \gamma \wedge \gamma \wedge \gamma + 9 \, \gamma \wedge \Gamma
\label{valdemoro}
\ee
where the index stands for Wang and Cassing who introduced this approximation in 1985 \cite{Wang}. 
This properly reduces to Eq.(\ref{NIA_app}) when we assume $\Gamma= 2 \gamma \wedge \gamma$.
\end{enumerate}

Now, if we want to go beyond these approximations, $\Delta^{(3)}$ in Eq.(\ref{cumu2-wedge}) must be 
approximated. In fact, there exist some approximations \cite{Nakatsuji, Mazziotti_approx, Mazziotti_CPL} 
-- for instance, by using the connection between RDMs and the Green's functions
\cite{Nakatsuji} -- but we are not going to deal with them in this paper.   

\section{Discussion Of The Results}
\label{sec_res}
In order to test the aforementioned approximations, we need a system for which we have access to its exact or nearly exact solution. 
The Hubbard model fits very well here since first of all we can solve it exactly for a few sites;
and secondly, since the number of single-particle orbitals that build the many-body Hilbert space is limited, 
we can retain the full single-particle basis set and avoid basis set truncation errors. It is worth mentioning that although 
here we illustrate only the performance of all the approximations for the Hubbard model, our preliminary
results for small molecules also exhibit very similar issues (work in progress \cite{javad}). 

Now, we consider the case of a linear finite Hubbard chain with only nearest neighbor interactions.
The Hubbard Hamiltonian for a finite one-dimensional system in second quantization notation is 

\be
\hat{H} \, = \, \sum_{\sigma, i} t \, (a_{i+1, \sigma}^{\dagger} a_{i,\sigma} + a_{i,\sigma}^{\dagger} a_{i+1,\sigma}) + 
\sum_i U \, n_{i\uparrow} n_{i\downarrow}
\label{Hubbard}
\ee
where $\sigma$ is a spin index, $i$ is the site index and $t$ and $U$ denote hopping and on-site Coulomb energy, respectively. Here, $t$ is set to unity and $U$ 
gets different values to simulate different correlation strengths. 

To study the quality of the approximations, we must go beyond two-particle systems since they can be treated exactly in our formalism. 
In this work, however, we will avoid the practical complications introduced by spin in odd-number-electron systems, and perform all our calculations 
in four electrons in a four-site system but that does not affect the generality of our results. A more detailed discussion on numerical
aspects is given in Appendix \ref{hubbard_eqs} and the code is also available upon request \cite{code}. 

In this part, we investigate three different approximations of $\Gamma^{(3)}$, namely the three-body collision integral free, the three-body non-interacting,
 and the WC approximations and compare them with the exact and the TDHF results.
With these approximations, we have now a closed set of equations and as any differential equation, we need an initial state of the system 
to propagate them. To study the initial state dependence of the phenomena, we choose two extreme regimes of initial states to perform our 
calculations: Far from equilibrium and near to equilibrium.

At first, we choose a far from equilibrium state as our initial state since it helps us to 
show the problem more clearly. We build such an initial state by putting four electrons in the two leftmost sites, \textit{i.e}.
\be
| \Psi_0 \rangle = a^\dagger_{1,\uparrow} a^\dagger_{1,\downarrow} a^\dagger_{2,\uparrow} a^\dagger_{2,\downarrow} | 0 \rangle 
\ee
where $1$ and $2$ refer to two neighboring sites at the beginning of the chain. In Appendix \ref{hubbard_eqs}, we give the matrix form of the 
initial state and the Eqs.(\ref{1rdm_BBGKY}) and (\ref{2rdm_BBGKY}) that we actually propagate.    

Here, the equations in hand are ordinary differential equations and we solve them numerically using the Runge-Kutta method. 
However, to ensure the accuracy and stability of our results, 
we also used more accurate time-propagation schemes such as the fourth-order Adams-Bashforth-Moulton method and we found no notable difference (for a detailed discussion of these 
methods see \cite{shampine}).

\begin{figure}
  \begin{center}
  \includegraphics[width=0.75\textwidth]{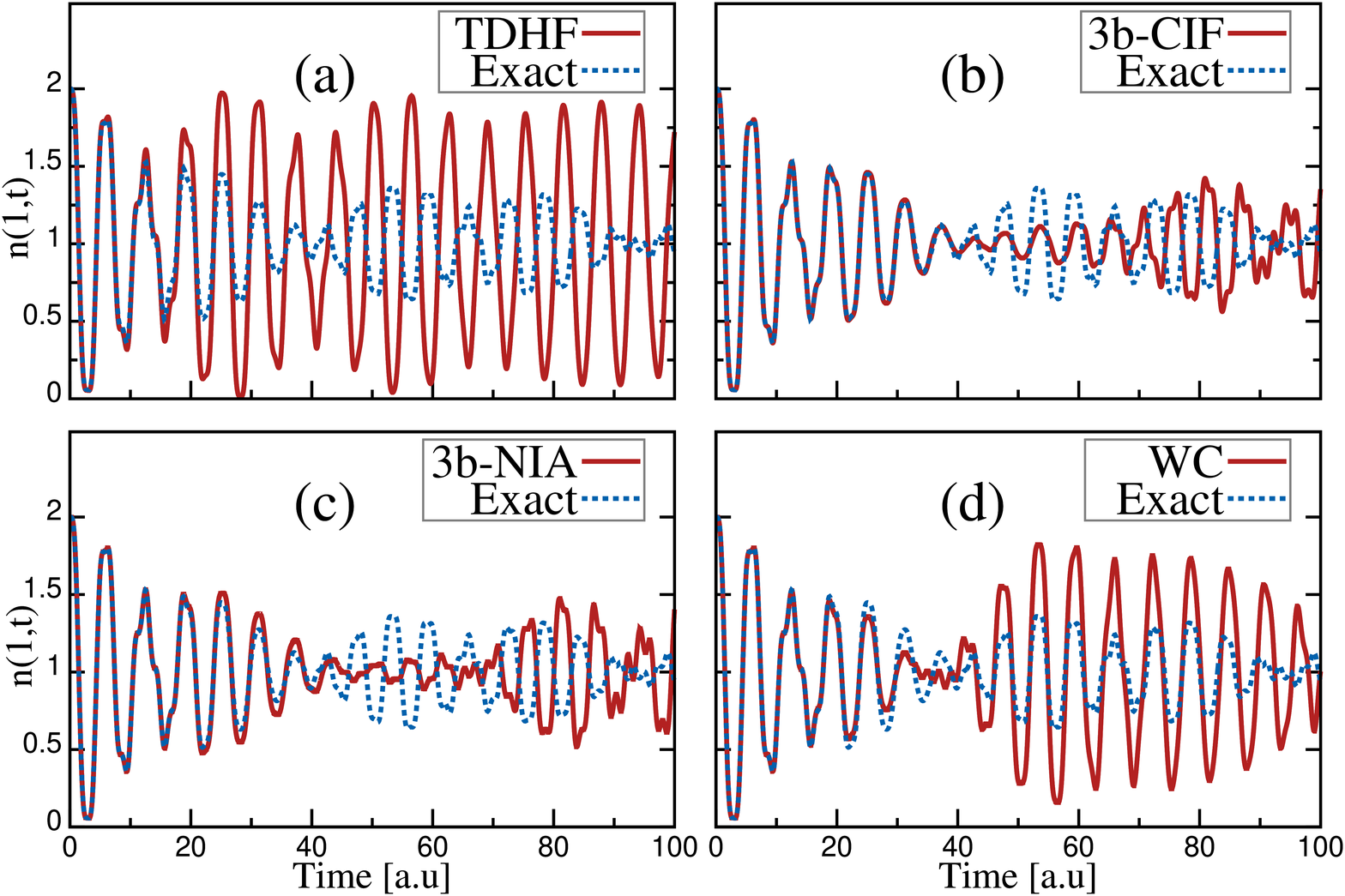}
  \caption{Time evolution of electronic density in the leftmost site of a 4-site Hubbard model with (a) TDHF   
               (b) 3b-CIF (c) 3b-NIA and, (d) WC approximations. The exact result is also 
               given for comparison. Here, $m$, $\hbar$ are set to
               unity and Hubbard parameters are $U=0.1$ and $t=1$.
               The four electrons filled the two leftmost sites initially.}
  \label{fig1}
  \end{center}
\end{figure}

The time evolution of electronic density in the leftmost site, n(1,t), is plotted in Fig.\ref{fig1} for a weak 
on-site Coulomb energy, $U=0.1$, and for (a) TDHF, (b) 3b-CIF, (c) 3b-NIA 
and, (d) WC approximation. The plots also contain the exact result for comparison. In a short-time 
scale, we can see that all three approximations improve the quality of the results considerably, compared 
to the TDHF. However, comparing with each other, the approximations do not exhibit large differences.
      
\begin{figure}
  \centering
  \includegraphics[height=0.6\textheight]{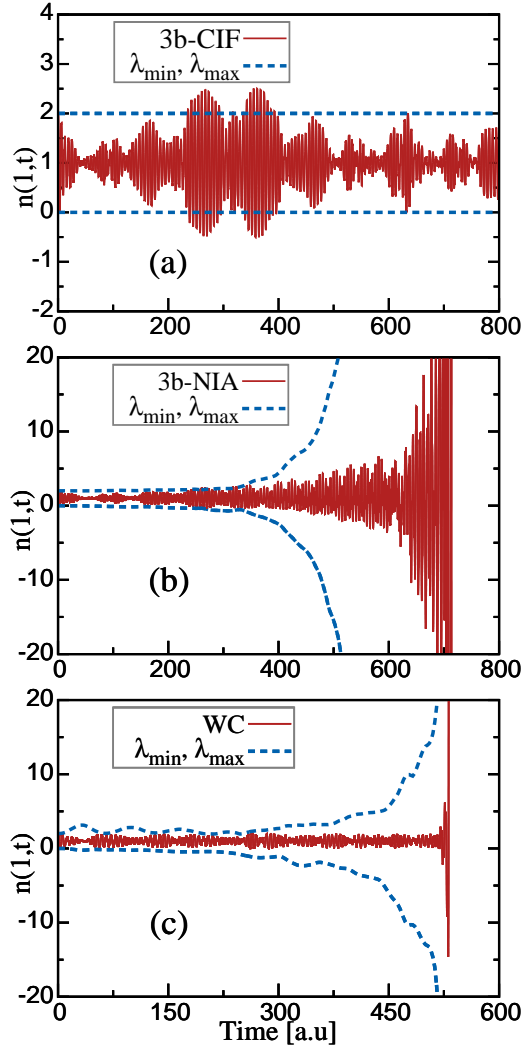} 
  \caption{Time evolution of electronic density in the leftmost site of a 4-site Hubbard model in a longer time scale for   
               (a) 3b-CIF (b) 3b-NIA, (c) WC approximations. Blue lines show the highest and lowest geminal occupation number
					in time. Here, $m$, $\hbar$ are set to
               unity and Hubbard parameters are $U=0.1$ and $t=1$.
               The four electrons filled the two leftmost sites initially.}
  \label{fig2}
\end{figure}

Figure \ref{fig2} shows essentially the same results for a longer propagation time. It also shows how the highest and lowest  
geminal occupation numbers, $\lambda_{max}$ and $\lambda_{min}$, behave in time. For the 3b-CIF in panel (a)
we can see unphysical behavior around  $ t \approx 240 \, a.u$, where the density acquires negative values or 
rises beyond two electrons in a site. The problem is more serious for the two other approximations since for longer 
propagation times, the electronic density starts to oscillate with amplitudes much beyond physically allowed boundaries, 
and eventually diverges as is shown in Fig.2 (b and c). The divergence time depends on the correlation strength, 
namely on the value of $U$ in our model, and it decreases almost exponentially with increasing $U$. For example, for 
WC approximation, the divergence time changes from $t \approx 532 \, a.u$ for $U=0.1$ to $t \approx 3 \, a.u$ 
for $U=10$. It is important to note that in 3b-NIA and WC approximations, $\lambda_{max}$ and $\lambda_{min}$
start to diverge much earlier, although we can not immediately see the effect in neither natural orbital occupation numbers nor on-site electronic
densities.

Nonetheless, it is well-known that the time-evolution of a far from equilibrium state is generally very difficult to handle with 
any approximation, and particularly with the ground-state-tuned ones; hence, 
we change the initial states to be closer to the system's ground state in order to investigate the generality of this phenomenon.
First, we start the simulation with the initial $\gamma$ and $\Gamma$ extracted from the ground state of i) the exact solution 
and ii) the Hartree-Fock approximation; and let it propagate with all three different approximations. Although in these cases 
the electronic density for the 3b-CIF does not violate physical bounds, we still see the divergence for other two approximations. 

Moreover, we used the method introduced by Mazziotti \cite{Mazziotti_GS_PRL} to find the ground state associated with 3b-NIA and
WC approximations and then used it as the initial state. However, since the method \cite{Mazziotti_GS_PRL} is not totally convergent, 
the result is not a truly stationary state and even starting from such state does not bring stability to the equations and divergence 
appears again. We will discuss this issue in more details in our forthcoming paper \cite{javad}.

These tests show that the divergence problem is independent of the initial state and has to do with the nature of the approximated equations. 
It is worth emphasizing again that in all of these approximations 
the continuity condition, Eq.(\ref{continuity}), has not been violated and the total number of particles is always conserved. 
Nevertheless, the continuity equation does not guarantee that the electronic density in each state does not go below zero or beyond two.

As we mentioned in the introduction, the violation of fermionic inequality has also been observed for a different
system in nuclear physics \cite{Schmitt_springer,Gherega1993166}.
In fact, there are earlier works in the classical BBGKY theory in which they studied the effect of nonlinearity 
introduced by truncation of the hierarchy, and showed the existence of instability in these
coupled equations depending on the initial conditions of the system \cite{schmit_Mckean,krieg}. Other studies also
indicated that the classical collision integral can diverge \cite{Goldman, Dorfman}.   
Such catastrophic behaviors of these coupled equations pose a valid question that why even highly advanced approximations based
on the Green's function expansion not only fail to follow fundamental physical principles, but also lead to divergence, 
even though the total energy and number of particles are conserved. 

To analyze this phenomenon we center our attention to the basic properties of the BBGKY hierarchy and density matrices
to find out how they are affected by different approximations. As we already showed, the employed approximations break 
the compatibility between Eq.(\ref{1rdm_BBGKY}) and the approximated version of Eq.(\ref{2rdm_BBGKY}) and the partial 
trace relation (Eq.(\ref{trace_rel})) between $\Gamma$ and $\gamma$ does not hold any more. Schmitt \textit{et al.} 
\cite{Schmitt_springer} and Gherega \textit{et al.} \cite{Gherega1993166} claimed  this to be the main source of the problem. 

On the other hand, it is obvious that the positive-semidefiniteness of density matrices
has also been violated. This problem may arise for one of the following reasons:

\begin{enumerate}[i)]
\item If in Eq.(\ref{2rdm_BBGKY}) the approximation functional of $\Gamma^{(3)}$ is built in a way that $\Gamma$ does not necessarily stay
positive-semidefinite, even though the initial $\gamma$ and $\Gamma$ are positive-semidefinite (see cases C and D below). Therefore, regardless
of whether the partial trace relation between $\gamma$ and $\Gamma$ holds or not, there is no guarantee for $\gamma$ to be positive-semidefinite.  
 
\item If the approximation functional of $\Gamma^{(3)}$ is built in a way that the propagated $\Gamma$ does stay
positive-semidefinite (provided the initial $\gamma$ and $\Gamma$ are positive-semidefinite), but since the Eqs. (\ref{1rdm_BBGKY}) 
and (\ref{2rdm_BBGKY}) are not compatible and the relation between $\Gamma$ and $\gamma$ is ill-defined, the positive-semidefiniteness
will not necessarily pass to the $\gamma$ (see case A below).

\end{enumerate}
It is not easy to impose positive-semidefiniteness on $\Gamma^{(3)}$ and even if it has such property, since its trace relation with 
$\gamma$ and $\Gamma$ is broken this does not lead to the positive-semidefiniteness of $\gamma$ and $\Gamma$. However, we are
exploring variational approaches to impose this constraint during the time evolution, namely preventing the system not to be positive-semidefiniteness.  

Next, we use different test approximations to analyze the role of compatibility and positive-semidefiniteness in these 
unphysical results.     
\subsection{Retaining Positive-semidefiniteness but not Compatibility.}

For the simplest case, 3b-CIF, the compatibility between 
second and first equation of the hierarchy is obviously lost. At the same time, it is easy to show that
the time evolution of $\Gamma$ is unitary in the sense that the solution to the equation of motion is of the form
\be
\Gamma(X_2, X'_2,t) = \sum_j \lambda_j \Phi_j (X_2,t) \Phi^*_j (X'_2,t). 
\label{secEq_nosource_ans}
\ee
where $\Phi_j$ satisfies the Schr\"odinger equation 
\be
i \partial_t \Phi_j (X_2,t) = \hat{H}_{12} \Phi_j (X_2,t)
\ee
Inserting Eq.(\ref{secEq_nosource_ans}) into Eq.(\ref{eigenvalue}) leads to time-independent geminal occupation numbers 
and provided the initial $\Gamma$ is positive-semidefinite, it preserves this feature at all times (Fig.\ref{fig2}(a)).
Note that this argument holds independent of possible time dependence in the two-particle Hamiltonian, $\hat{H}_{12}$.

In addition, substituting Eq.(\ref{secEq_nosource_ans}) in the right-hand side of Eq.(\ref{1rdm_BBGKY}) yields a simple linear equation for
the time propagation of $\gamma(\bx_1,\bx'_1,t)$ which  does not diverge. 
Nevertheless, because the partial trace relation (Eq.(\ref{trace_rel})) between $\gamma$ and $\Gamma$ is no more valid, 
natural orbital occupation numbers may not necessarily lie between zero and one \cite{coleman} and as a consequence the density can gain
negative values (Fig.\ref{fig2}(a)). With this example we come to the conclusion that positive-semidefiniteness of the 
$\Gamma$ is not a sufficient condition for keeping the $\gamma$ positive-semidefinite.
 
\subsection{Retaining Compatibility and Positive-semidefiniteness.}

It is very insightful to look for an approximation that retains both of these properties and see how well that can perform in the calculation. 
The simplest possible approach is to replace the three-body collision integral term, $S$ in Eq.(\ref{2rdm_BBGKY}), with another term under the 
condition that the two equations become compatible. This condition, of course, does not determine a unique term, but to study its effect we 
consider the following simple expression
\be
S(X_2, X'_2,t) = (N-2) (U(X_2) - U(X'_2)) \, \Gamma(X_2, X'_2,t).
\label{compatibel}
\ee
Thus, the second equation of the hierarchy becomes
\bea
&& (i  \, \prt  - \hat{H}_{12} + \hat{H}_{1'2'} ) \, \Gamma(X_2, X'_2,t) = (N-2) (U(X_2) - U(X'_2)) \, \Gamma(X_2, X'_2,t).
\label{2rdm_BBGKY_comp}
\eea
When in above equation, we put $\bx_2=\bx_2'$ and integrate over $\bx_2$ we obtain the first equation of 
the BBGKY hierarchy, Eq.(\ref{1rdm_BBGKY}).
\begin{figure}
  \centering
   \includegraphics[width=0.75\textwidth]{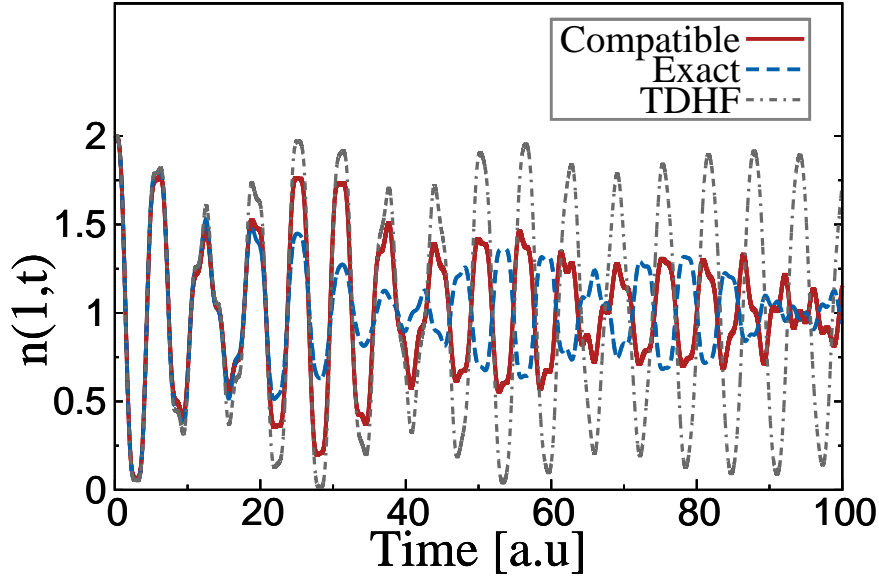} 
  \caption{Number of electrons in the leftmost site of a 4-site Hubbard model with compatible approximation. 
               The exact and TDHF results are also given for comparison.  
               Here, $m$, $\hbar$ are set to unity and Hubbard parameters are $U=0.1$ and $t=1$.
               The four electrons filled the two leftmost sites initially.}
  \label{fig3}
\end{figure}

This approximation makes Eq.(\ref{2rdm_BBGKY}) linear. Moreover, it retains the compatibility between two equations, meaning the two approaches for solving the BBGKY hierarchy, 
discussed in section \ref{EOM}, are equivalent. We call this the compatible approximation. The effect of this approximation is equivalent to magnifying the interparticle interaction 
by a factor of $(N-1)$ in the 3b-CIF approximation. Hence, we can modify 
interparticle interaction accordingly in Eq.(\ref{secEq_nosource_ans}) and use the same argument 
to prove that this approximation also keeps $\Gamma$ positive-semidefinite. Now, as a result of compatibility, $\gamma = \gamma_{_{\Gamma}}$ and
this $\gamma$ immediately inherits positive-semidefiniteness of $\Gamma$ and in this fashion, compatibility 
and positive-semidefiniteness are both incorporated. The results for this approximation are depicted in the Fig.(\ref{fig3}) for 
a limited time scale. As expected, we do not observe any violation in the electronic density even after long propagation.

Despite the well-behaved result, it should be mentioned that whereas the natural orbital occupation numbers cannot acquire negative values, they may still 
exceed one (the density at one point may develop beyond two particles). For example, we observed such violation for natural orbital occupation numbers in a tiny time 
region only when the initial state was far from equilibrium, \textit{i.e.} four electrons occupying the two leftmost sites, and $U$ was very large ($U=10$).

\subsection{Retaining Compatibility but not Positive-semidefiniteness.}
\label{compNOposit}

In order to find out the role of positive-semidefiniteness and its importance, in the next step, we build an approximation that fulfills 
the compatibility but not positive-semidefiniteness. To accomplish this goal, we add a term, $Z$, with a partial trace summing up to zero, to 
the previous compatible approximation, Eq.(\ref{compatibel}). Substituting this term in Eq.(\ref{2rdm_BBGKY}) we have  
\bea
&& (i \, \prt  - \hat{H}_{12} + \hat{H}_{1'2'} ) \, \Gamma(X_2, X'_2,t)  =  
\nonumber \\
&& \phantom{i \, \prt  - \hat{H}_{12} + } (N-2) (U(X_2) - U(X'_2)) \, \Gamma(X_2, X'_2,t) + Z(X_2, X'_2,t)
\label{2rdm_BBGKY_comp2}
\eea
where 
\be
\int d\bx_2 \, Z(\bx_1,\bx_2, \bx'_1,\bx_2,t)  \, = 0
\label{int_Z}
\ee
which in most cases, $Z$ is a nonlinear functional of density matrices. In this way, both levels of the hierarchy stay compatible but the additional term does not necessarily keep $\Gamma$ positive-semidefinite.  
This term is not unique and in fact it can be derived for different approximations (see Appendix \ref{zero_traced} for detailed derivation). 

Nonetheless, although these approximations retain the compatibility, they do not necessarily keep positive-semidefiniteness of RDMs and they 
eventually lead to divergence if we propagate them long enough in time. These results are the evidence that compatibility is not a sufficient 
condition to keep the equations bounded.

\subsection{Using Only the Second Equation (Eq.(\ref{2rdm_BBGKY})). }
\label{sec_only}

\begin{figure}
  \centering
   \includegraphics[width=0.75\textwidth]{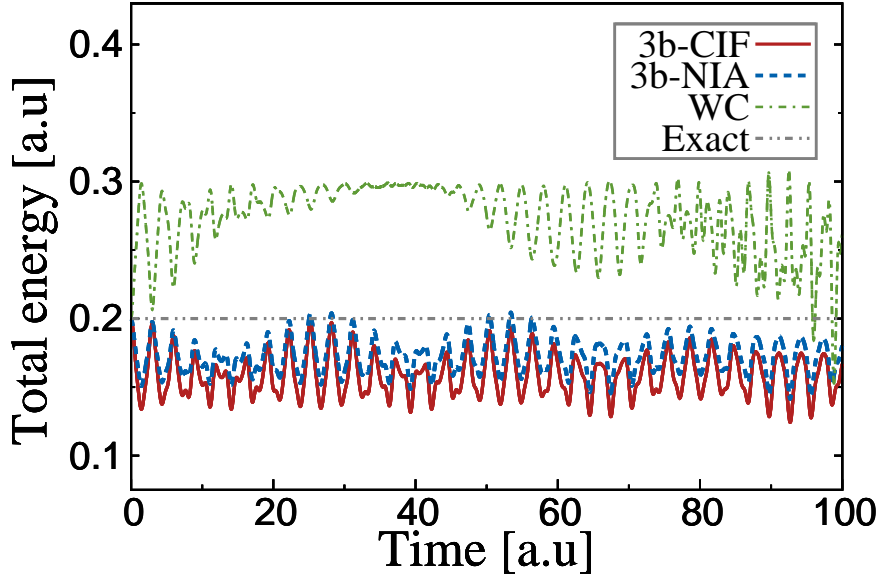} 
  \caption{Fluctuations of total energy of the 4-site Hubbard model for different approximations using only Eq.(\ref{2rdm_BBGKY}). 
               The exact result is also given for comparison.   
               Here, $m$ and $\hbar$ are set to unity and Hubbard parameters are $U=0.1$ and $t=1$.
               Four electrons filled the two leftmost sites initially.}
  \label{fig4}
\end{figure}

As we discussed in section \ref{EOM} in details, there are two approaches to solve the BBGKY hierarchy in the second level. So far in all discussed
approximations, we have propagated both Eqs.(\ref{1rdm_BBGKY}) and (\ref{2rdm_BBGKY}) together; now, we turn to the second approach and solve only 
Eq.(\ref{2rdm_BBGKY}) for above approximations. However, as we showed earlier, the total energy will not be necessarily conserved in this method unless 
the approximation makes the two equations compatible. Figure \ref{fig4} shows that the energy fluctuations in time are relatively large for all the approximations,
which puts a question mark over the quality of this approach in general. 

Nonetheless in this way, the compatibility between equations is not an issue anymore which helps us to study the behavior 
of only the second equation. Our calculations show that, even though the fermionic inequality violation 
of 3b-CIF approximation has been cured in this way, the divergence in 3b-NIA or WC approximations has surfaced again and in fact, 
even the divergence time has not been improved in any of them. With these observations, we can confidently claim that the incompatibility is not 
the only cause of diverging behavior of the equations, and the role of positive-semidefiniteness should also be taken into account. 

In these four subsections, we provided many test approximations fulfilling different constrains to show that neither compatibility between equations nor 
positive-semidefiniteness of the approximations by itself, can keep the propagation of the RDMs inside the fermionic boundaries. In fact, although 
the nonlinearity introduced by most of the approximations to Eq.(\ref{2rdm_BBGKY}), might be the cause of divergence, 
the violation of fermionic inequality might exist even in the case of linear approximations as we saw in the 3b-CIF approximation.
Therefore, it indeed takes both of these approximation constrains to tame such coupled equations.

\section{Summary and Conclusions}
\label{conclusion}

In this work, we pointed out the main challenges we face in order to decouple the hierarchy of equations for the time evolution of density matrices.
In particular, we studied several approximations for the three-body RDM in terms of the one- and two-body RDMs. 
First, we showed that once an approximation
for the three-body RDM is made, the equation of motion for the two-body RDM does no longer imply 
the validity of the first equation of the hierarchy. Therefore, in order to obey energy conservation it is necessary to solve the equations of motion for 
the one-body and two-body density matrices simultaneously. 

Next, we studied numerical solutions for
the fermionic Hubbard model for several of the decoupling schemes and compared them to exact results obtained from solving the time-dependent
Schr\"odinger equation. We found that in many decoupling schemes the local electron density attains unphysical negative values in time and 
natural orbital occupation numbers in general do not remain between zero and one as is required for fermionic systems.
Furthermore, in most of existing approximations the local electron density diverges, although total particle number and energy are perfectly conserved. 
We investigated whether this feature
is cured by forcing the two lowest equations of the hierarchy to be compatible and found this not to be the case.

We conclude that, a possible way to make progress in the application of the BBGKY hierarchy is to make sure that positive-semidefiniteness of the density matrices and the
fermionic constraints on the occupation numbers is built into the equations. 
For example in the case of the time-dependent Hartree-Fock, there exists always a wave function corresponding to the first
equation of the hierarchy and therefore the natural orbital occupation numbers can never be unphysical. 
This suggests a further study of the BBGKY equations in relation to (approximate) wave functions. 
Further progress can be made in linear response theory
since in the linear response regime the nonlinearities are, by definition, removed and the BBGKY approach can be
investigated further for the study of optical spectra. Work along these lines is in progress. 

\section{Acknowledgment}
The authors wish to thank Professor E. K. U Gross for useful discussions. We are also
grateful to Professor Ilya Tokatly and  Ari Harju for their comments on the manuscript.  
We acknowledge financial support from the European Research Council Advanced Grant DYNamo (ERC-2010-AdG, Proposal No. 267374),
Spanish Grants (FIS2011-65702-C02-01 and PIB2010US-00652), ACI-Promociona (ACI2009-1036), 
Grupos Consolidados UPV/EHU del Gobierno Vasco (IT-319-07), Consolider nanoTHERM (Grant No. CSD2010-00044), 
Fondo Social Europeo (FSE) and Spanish Research Council (CSIC) through its "Junta para la Ampliaci\'{o}n de Estudios" fellowship
(JAE-Predoc-2008), European Commission projects CRONOS (280879-2 CRONOS CP-FP7) and THEMA(FP7-NMP-2008-SMALL-2, 228539). 
We are also grateful to the support of Academy of Finland through its "Center of Excellence" Grant (2006-2011).

\appendix

\section{Energy Conservation Derivation}
\label{appen-energy}

Now we discuss in full details the energy conservation in the BBGKY hierarchy. According to 
Eqs.(\ref{e1}) and (\ref{e2}), the change in the total energy of the system, $E(t)=E_1 (t) + E_2 (t)$, 
is given by
\be
\frac{dE}{dt}= \frac{dE_1}{dt} + \frac{dE_2}{dt}
\ee
where
\bea
\frac{dE_1}{dt} &=&  \int d\bx' \, \partial_t v(\bx',t) \gamma (\bx,\bx',t) |_{\bx=\bx'} + \int d\bx' \, h(\bx',t) \partial_t 
\gamma (\bx,\bx',t) |_{\bx=\bx'}
\label{e1t}
\eea
and
\bea
\frac{dE_2}{dt} &=&  \frac{1}{2} \int d\bx_1 d\bx_2 \, U (X_2) \partial_t \Gamma (X_2 ,X_2,t).
\label{e2t}
\eea
To proceed, we replace the time dervative of $\Gamma$ from Eq.(\ref{2rdm_BBGKY}), however for
$\gamma$ in Eq.(\ref{e1t}), we can use either Eq.(\ref{1rdm_BBGKY}) or (\ref{1rdm_BBGKY2}). In 
our derivation, we use Eq.(\ref{1rdm_BBGKY2}) and the case of Eq.(\ref{1rdm_BBGKY}) is obtained as the special case by
taking $\bar{\Gamma}=\Gamma$ at the end of the derivation.
Let us start by evaluating (\ref{e2t}). The equation of motion (\ref{2rdm_BBGKY}) yields
\bea
\frac{dE_2}{dt} &=& - \frac{1}{4i} \int d\bx_1 d\bx_2 \, U (X_2) (\nabla_1^2 + \nabla_2^2- \nabla_{1'}^{2}-\nabla_{2'}^{2})
\Gamma (X_2 ,X'_2,t) |_{1,2=1',2'} \nonumber \\
&=& - \frac{1}{2i} \int d\bx_1 d\bx_2 \, U (\bx_1 \bx_2) (\nabla_1^2 - \nabla_{1'}^{2})
\Gamma (\bx_1 \bx_2 , \bx_1' \bx_2,t) |_{1=1'} \nonumber \\
&=& \frac{1}{2i} \int d\bx_1 \, \nabla_1 U (\bx_1 \bx_2 ) \cdot \left[ ( \nabla_1 -\nabla_{1'}) \Gamma (\bx_1 \bx_2 , \bx_1' \bx_2,t) |_{1=1'} \right]
\eea
where we used the symmetry of $\Gamma$ and performed a partial integration.
Now, we consider the change in the one-body energy, $E_1 (t)$, primarily by evaluating the second term in Eq.(\ref{e1t}).
It reads
\bea
\int d\bx \, h(\bx',t) \partial_t 
\gamma_{_{\Gamma}} (\bx,\bx',t) |_{\bx=\bx'} = \int d\bx \, ( v(\bx,t) \partial_t n_{_{\Gamma}}(\bx,t) + \frac{1}{2} \partial_t \nabla\cdot \nabla' \gamma_{_{\Gamma}} (\bx,\bx',t)|_{\bx=\bx'})
\label{step1b}
\eea
where in the second term of the right-hand side, we used 
\be
\int d\bx d\bx' \delta (\bx-\bx') \nabla^{'2} \gamma_{_{\Gamma}} (\bx,\bx',t) = - \int d\bx d\bx' \delta (\bx-\bx')  \nabla\cdot \nabla' 
\gamma_{_{\Gamma}} (\bx,\bx',t).
\ee
On the other hand, from the equation of motion (\ref{1rdm_BBGKY2}) it follows 
\bea
\label{td_nabla}
i \partial_t \nabla_1 \cdot \nabla_{1'} \gamma_{_{\Gamma}} (\bx_1,\bx_{1'},t)|_{1=1'} &=& (\nabla_1 v(\bx_1,t) \cdot \nabla_{1'} - \nabla_{1'} \, v(\bx_1',t) \cdot \nabla_1)
\gamma_{_{\Gamma}} (\bx_1,\bx_{1'},t)|_{1=1'}  \\
&& - \frac{1}{2} \nabla_1 \cdot 
\left[ ( \nabla - \nabla' ) (\nabla_1 \cdot \nabla_{1'})\gamma_{_{\Gamma}}(\bx,\bx',t)  \bigr|_{1'=1'} \right] \nonumber \\
&& + (\nabla_1 \cdot \nabla_{1'}) 
\int d\bx_2 \, (U(\bx_1\bx_2) - U(\bx'_1\bx_2)) \bar{\Gamma}(\bx_1 \bx_2,\bx'_1\bx_2,t) |_{1=1'}. \nonumber 
\eea
The second term in the right-hand side of this expression is a total derivative and vanishes upon integration. 
Putting Eq.(\ref{td_nabla}) back into (\ref{step1b}) and the resulted expression in Eq.(\ref{e1t}), we get
\bea
\frac{dE_1}{dt} &=&  \int d\bx \, ( \partial_t v(\bx,t) n_{_{\Gamma}}(\bx,t) + v(\bx,t) (\partial_t n_{_{\Gamma}}(\bx,t)+ \nabla \cdot
 \mathbf{j}_{_{\Gamma}} (\bx,t)) )\nonumber \\
&& + \frac{1}{2i} \int d\bx_1 d\bx_2 (\nabla_1 U (\bx_1 \bx_2) \cdot \nabla_{1'} - \nabla_{1'} U(\bx_1'\bx_2) \cdot \nabla_{1})
  \bar{\Gamma}(\bx_1 \bx_2,\bx'_1\bx_2,t) |_{1=1'} \, .
\eea
The second term under the first integral vanishes since the continuity equation holds. Therefore, for the total energy
 we finally arrive at 
\bea
\label{energy_conv_app}
\frac{dE}{dt} &=& \int d \bx \, \partial_t v(\bx,t) n_{_{\Gamma}}(\bx,t) \\
&& + \frac{1}{2i} \int d\bx_1 d\bx_2 \, \nabla_1 
U (\bx_1 \bx_2 ) \cdot \left[ ( \nabla_1 -\nabla_{1'}) (\Gamma (\bx_1 \bx_2 , \bx_1' \bx_2,t)-\bar{\Gamma} 
(\bx_1 \bx_2 , \bx_1' \bx_2,t))  |_{1=1'} \right]. \nonumber 
\eea

In the case that $\gamma$ evolves through Eq.(\ref{1rdm_BBGKY}), we must replace $\bar{\Gamma}$ with 
$\Gamma$ and hence the final result reads
\be
\label{energy_conv_first_approach}
\frac{dE}{dt} = \int d \bx \, \partial_t v(\bx,t) n(\bx,t)
\ee
which is exactly the energy conservation law as it is discussed in the main text.

\section{Computational Details for the Hubbard Model}
\label{hubbard_eqs}
The real space description of Eqs.(\ref{1rdm_BBGKY}) and Eq.(\ref{2rdm_BBGKY}) makes them very impractical for computational purposes.
Therefore, they must be transformed to the matrix from, using an appropriate basis set.
In the case of Hubbard model, the basis set is made of site-orbitals for each spin; for example, for a four-site Hubbard system, the basis set
contains eight orbitals. 

Here we only concern with spin-compensated systems(\textit{i.e.} $S_z = S =0$), in which we can use many symmetries to
eliminate the spins and simplify the equations of motion \cite{McWeeny}. We define

\be
\gamma (\br_1 \br_1',t) = \sum_{\sigma_1} \gamma(\br_1\sigma_1 , \br_1'\sigma_1,t )
\ee
and
\bea
\Gamma (\br_1 \br_2, \br_1' \br_2',t) =
 \sum_{\sigma_1 \sigma_2} \, \Gamma(\br_1 \sigma_1 \br_2\sigma_2, \br_1' \sigma_1 \br_2'\sigma_2,t).
\eea
where $\sigma$ denotes the spin coordinate.
Therefore, the matrix from of Eqs.(\ref{1rdm_BBGKY}) and (\ref{2rdm_BBGKY}) for a $M$-site Hubbard model in site-orbital basis set read     
\bea
i \prt \gamma_{ij} & = & t \left(\gamma_{i\,(j+1)} (1-\delta_{jM}) + \gamma_{i\,(j-1)} (1-\delta_{j1}) 
-\gamma_{(i+1)\,j} (1-\delta_{iM}) - \gamma_{(i-1)\,j} (1-\delta_{i1}) \right)  \nonumber \\
&& + \, U \left( \Gamma_{i\,j\,j\,j} -\Gamma_{i\,i\,j\,i} \right)
\label{eq:gammamtrix}
\eea
and
\bea
\label{eq:Gammamatrix}
&& i \prt  \Gamma_{ijkl} = t \bigl( \Gamma_{ij(k+1)l} (1-\delta_{kM})
+ \Gamma_{ij(k+1)l} (1-\delta_{k1}) + \Gamma_{ijk(l+1)} (1-\delta_{lM})
+ \Gamma_{ijk(l-1)} (1-\delta_{l1}) \nonumber \\ 
&& \phantom{ \prt  \Gamma_{ijkl}  = t \bigl(} - \, \Gamma_{(i+1)jkl} (1-\delta_{iM})
- \Gamma_{(i-1)jkl} (1-\delta_{i1}) - \Gamma_{i(j+1)kl} (1-\delta_{jM})
 - \Gamma_{i(j-1)kl} (1-\delta_{j1}) \bigr) \nonumber \\ 
&& \phantom{ \prt  \Gamma_{ijkl}  = } +  U \bigl( \Gamma_{ijkk} \, \delta_{kl} - \Gamma_{iikk} \, \delta_{ij}  \bigr)
\nonumber \\
&&\phantom{ \prt  \Gamma_{ijkl} = }  + U  \sum_{\sigma_i \sigma_j \sigma_n} \bigl( \Gamma^{(3)^{\sigma_i \sigma_j \sigma_n \sigma_i \sigma_j \sigma_n}}_{\quad i\,j\,k\,k\,l\,k} 
+ \Gamma^{(3)^{\sigma_i \sigma_j \sigma_n \sigma_i \sigma_j \sigma_n}}_{\quad i\,j\,l\,k\,l\,l} 
- \Gamma^{(3)^{\sigma_i \sigma_j \sigma_n \sigma_i \sigma_j \sigma_n}}_{\quad i\,j\,i\,k\,l\,i} 
- \Gamma^{(3)^{\sigma_i \sigma_j \sigma_n \sigma_i \sigma_j \sigma_n}}_{\quad i\,j\,j\,k\,l\,j} \bigr) 
\eea
where indices denote the site numbers. The $\Gamma^{(3)}$ should be replaced by an approximation and then the spins can be integrated out using
the same symmetry argument. These are the actual equations that we solve in this work. 

As it is mentioned in the main text, we choose a four-site Hubbard model where four electrons initially filled the two leftmost sites so that
\be
\gamma_{i,j} = 
\begin{cases}
\gamma_{i,i}=2, &  i=1,2   \\
\gamma_{i,j}=0, & \text{for the rest.}   
\end{cases}
\ee
Now, since this initial state is a Slater determinant formed by two site-orbitals, $\Gamma$ has the exact 
form of Eq.(\ref{g2_HF}) and can be written as
\be
\label{GammaFromgamma}
\Gamma_{ijkl} =  \gamma_{i,k} \gamma_{j,l} -\frac{1}{2} \gamma_{i,l} \gamma_{j,k}.
\ee
For the initial state being the ground state of HF, we use Eq.(\ref{GammaFromgamma}) with the same argument and construct the $\Gamma$ from $\gamma$.
Evidently, we cannot use the same argument for any initial state.
In the case of starting from the exact ground state, we extract the exact $\gamma$ and $\Gamma$ and feed them into the equations.

\section{Deriving The Zero-Traced Term}
\label{zero_traced}
Here, we derive the zero-traced expression, $Z$, which we introduced in section \ref{compNOposit}. This additional 
term can be written for different $\Gamma^{(3)}$ approximations as
\bea
Z(X_2,X'_2, t) & = & 
 \int d\bx_3 \,(U(\bx_1\bx_3) + U(\bx_2\bx_3)- U(\bx'_1\bx_3)- U(\bx'_2\bx_3)) \, 
\Gamma^{(3)}_{app} (\bx_1 \bx_2 \bx_3,\bx'_1\bx'_2 \bx_3,t) \nonumber \\
&& - F_{app}(\gamma,\Gamma)  
\label{additional_term }
\eea
where the first term is the 3-body collision integral with $\Gamma^{(3)}$ substituted by an approximation
and $F_{app}(\gamma,\Gamma)$ is defined for that approximation as
\be
Tr_{\bx_2}(F_{app}(\gamma,\Gamma)) = 
\int d\bx_3 \,d\bx_2 \,(U(\bx_1\bx_3) - U(\bx'_1\bx_3)) 
\, \Gamma^{(3)}_{app} (\bx_1 \bx_2 \bx_3,\bx'_1\bx_2 \bx_3,t).
\label{F_gG}
\ee
Evidently the partial trace of $Z$ will be zero.
 
As an example, we give the explicit form of $F_{3b-NIA}(\gamma,\Gamma)$ as
\bea
&& F_{3b-NIA}(\gamma,\Gamma)  =   (U(\bx_1 \bx_2) - U(\bx'_1 \bx'_2)) \, \bigl \{ 
N \, (\gamma(\bx_1,\bx'_1) \gamma(\bx_2,\bx'_2) - \gamma(\bx_1,\bx'_2) \gamma(\bx_2,\bx'_1))
\nonumber \\ 
&& \phantom{F(\gamma,\Gamma)  =  (U(\bx_1 \bx_2) - U(\bx'_1 \bx'_2)) \quad  }
+ g(\bx_1,\bx'_2)\gamma(\bx_2,\bx'_1)- g(\bx_2,\bx'_2)\gamma(\bx_1,\bx'_1) 
\nonumber \\
&& \phantom{F(\gamma,\Gamma)  =  (U(\bx_1 \bx_2) - U(\bx'_1 \bx'_2)) \quad  }
+ g(\bx_2,\bx'_1)\gamma(\bx_1,\bx'_2)- g(\bx_1,\bx'_1)\gamma(\bx_2,\bx'_2) \bigr\}
\label{HF_compatible}
\eea
where 
\be
g(\bx'_1,\bx_1) = \int \, d\bx_2 \gamma(\bx'_1,\bx_2) \gamma(\bx_2,\bx_1).
\label{g_def}
\ee
In the same way, $F(\gamma,\Gamma)$ can be obtained for other approximations but
we do not present them here. Therefore, the full expression to replace the 3-body collision integral reads
\bea
S(X_2,X'_2,t) & = & (N-2) (U(X_2) - U(X'_2)) \, \Gamma(X_2,X'_2,t) \\ 
&& + \int d\bx_3 \,(U(\bx_1\bx_3) + U(\bx_2\bx_3)- U(\bx'_1\bx_3)- U(\bx'_2\bx_3))
\, \Gamma^{(3)}_{app} (\bx_1 \bx_2 \bx_3,\bx'_1\bx'_2 \bx_3,t) \nonumber \\ 
&& - F_{app}(\gamma,\Gamma)  \nonumber 
\label{making_compatible}
\eea
where the first term is coming from the Eq.(\ref{compatibel}).


%

\end{document}